\documentclass[aps,superscriptaddress,reprint,onecolumn,11pt,nofootinbib]{revtex4-2}
\usepackage[english]{babel}
\usepackage[utf8]{inputenc}
\usepackage[T1]{fontenc}
\usepackage{enumitem,commath,caption,subcaption}
\usepackage{amsmath,graphicx}
\usepackage[export]{adjustbox}
\usepackage[colorlinks=true,allcolors=blue]{hyperref}
\usepackage{color,booktabs,multirow}
\usepackage[normalem]{ulem}
\begin{document}
\captionsetup{justification=raggedright,singlelinecheck=false,font=small}
%%%%%%%%%%%%%%%%%%%%%%%%%%%
\title{Optically dense nanowire metamaterials are transparent to polarization}

%%%%%%%%%%%%%%%%%%%%%%%%%%%
\title{Optically dense nanowire metamaterials are transparent to polarization}

\author{Shravan Raghunathan}
\affiliation{Complex Photonic Systems (COPS) Chair, Department of Science and Technology, University of Twente, P.O. Box 217, 7500 AE Enschede, The Netherlands}
\affiliation{Present address: Leibniz Institute of Photonic Technology (IPHT), Albert-Einstein-Straße 9, 07745 Jena, Germany}

\author{Ad Lagendijk}
\affiliation{Complex Photonic Systems (COPS) Chair, Department of Science and Technology, University of Twente, P.O. Box 217, 7500 AE Enschede, The Netherlands}
\affiliation{Present address: Complex Photonic Systems (COPS) group, Photonic and Semiconductor Nanostructures (PSN) Chair, Department of Applied Physics and Science Education (APSE), Eindhoven University of Technology (TU/e), P.O. Box 513, 5600 MB Eindhoven, The Netherlands}

\author{Willem L. Vos}
\thanks{Corresponding author: w.l.vos@tue.nl}
\affiliation{Complex Photonic Systems (COPS) Chair, Department of Science and Technology, University of Twente, P.O. Box 217, 7500 AE Enschede, The Netherlands}
\affiliation{Present address: Complex Photonic Systems (COPS) group, Photonic and Semiconductor Nanostructures (PSN) Chair, Department of Applied Physics and Science Education (APSE), Eindhoven University of Technology (TU/e), P.O. Box 513, 5600 MB Eindhoven, The Netherlands}

%%%%%%%%%%%%%%%%%%%%%%%%%%%
\date{\today}
%%%%%%%%%%%%%%%%%%%%%%%%%%%

\begin{abstract} 
We study the transport of light through dense opaque anisotropic metamaterials consisting of oriented nanowires. 
The nanowires consist of polymer photoresist that is structured by direct laser writing (DLW) with two-photon induced polymerization, with radii between $a = 0.5$ and $1~\mu \text{m}$. 
Our flat samples have a thickness up to 9 layers, from $L = 3~\mu \text{m}$ to $20~\mu \text{m}$. 
Within each layer, the nanowires are parallel and spaced with random nearest-neighbor distances; nanowires in adjacent layers are perpendicular. 
The diffuse optical transmission at $\lambda = 633~$nm is as low as $T = 12 \%$, typical of optically dense, multiple scattering metamaterials, with a mean free path down to $\ell = 1.1~\mu \text{m}$, much less than the sample thickness.
It is striking that the linear polarization of the input light is maintained at the output of the dense nanowire samples, and not scrambled as in dense nanosphere arrays. 
Moreover, the linear output polarization faithfully tracks the input polarization. 
We propose that the polarization is maintained in our optically thick samples, since light is predominantly transported perpendicularly to the nanowire layers. 
The polarization vector then lies in the nanowire plane, consisting of a linear combination of parallel and perpendicular vectors that are both conserved upon subsequent scattering. 
Hence, the polarization remains independent of nanowire orientation, even after multiple scattering events.
We propose that anisotropic scattering samples may find practical uses in white LEDs and its applications in lighting luminaires, optical communication, and encryption systems. 
\end{abstract}
%%%%%%%%%%%%%%%%%%%%%%%%%%
%%%%%%%%%%%%%%%%%%%%%%%%%% body

\maketitle

\section{Introduction} \label{sec:introduction} 
%%%%%%%%%%%%%%%%%%%%%%%%%%
The study of light transport in complex inhomogeneous metamaterials such as biological tissue, foam, or paint, is enjoying a considerable worldwide attention~\cite{Lagendijk1996PhysRep, vanRossum1999RMP, Akkermans2007Book, Carminati2021book}. 
While this field of 'complex photonics' started in fundamental physics, inspired by the analogies between the intriguing interference phenomena of electromagnetic waves that multiply scatter in complex materials and electrons scattering inside mesoscopic nanostructures~\cite{vanHaeringen1990book, Lagendijk1996PhysRep, vanRossum1999RMP, Akkermans2007Book}, it has recently been realized that major industrial applications profit from the advanced know-how of light scattering, ranging from semiconductor metrology for computers and smart phones~\cite{Orji2018ne}, via light emitting diodes for home, car, and street lighting~\cite{Schubert2006Book, Akasaki2015RMP, Amano2015RMP, Nakamura2015RMP, Meretska2019ACSPhot}, to satellite optics for environmental and climate monitoring~\cite{vos2016FFSO}. 
The transport of light through opaque metamaterials is in essence determined by the scattering of the polarized electric fields by the light scattering building blocks (`scatterers') inside the medium. 
In the elastic scattering regime where absorption is negligible, the incident light waves are multiply scattered, hence the direction of the light becomes scrambled inside the medium, compared to the incident direction. 
In opaque scattering samples, such as paint layers, tissue, or foam, the scrambling of the directionality is commonly accompanied by a complete scrambling of the polarization of the incident light. 

The majority of light transport studies in three dimensional (3D) complex media invoke systems that are composed of spherical nanoparticles as building blocks~\cite{Akdemir2024PRA}. 
While it is well-known that scattering of light by spheres is anisotropic - on account of light being a vector wave~\cite{Bohren1998book, Goldstein2003book} - a broad range of new applications are envisioned if the scattering building blocks themselves have anisotropic shapes, for instance, ellipsoids, cylinders~\cite{Lee98josaa, lee2008josaa,Huang2019ijmmct}, nanopores~\cite{Huisman2012PRL}, scatterers with an anisotropic dielectric function~\cite{Wiersma1999PRL, Wiersma2000PRE}, or even scatterers embedded in an anisotropic medium~\cite{Kaas2008PRL}. 
Relevant applications vary from the study of biological tissue~\cite{Guo2016jbo, Sun2018ol, parker2019pmb}, or of atmospheric dust~\cite{potenza2016sr, Jaiswal2021pss}, to nematic liquid crystals~\cite{vanTiggelen1996prl, Stark1996prl, vanTiggelen2000rmp, Sapienza2004prl, Choudhary2025SciRep}. 
When anisotropic random media interact strongly with light, 3D Anderson localization of light - where wave transport is completely halted by interference - has been predicted to occur more readily than in isotropic materials~\cite{Zhang1990PRB, Zambetaki1996PRL}.

In this paper, we perform a study of the transport of polarized light in oriented ensembles of nanowires as basic scattering units, that are made by direct laser writing (DLW) using two-photon polymerization~\cite{buckmann2012am}. 
This fabrication method allows us to finely control both the shape and the orientation of the scattering nanowires. 

%%%%%%%%%%%%%%%%%
\section{Experimental}\label{sec:experimental}
%%%%%%%%%%%%%%%%%
%%%%%%%%%%%%%%%%% 
\subsection{Samples}\label{sec:samples}
%%%%%%%%%%%%%%%%%

%%%%%%%%%%%%%%%%%
\begin{figure}[h!]
\centering\includegraphics[width=0.75\columnwidth]{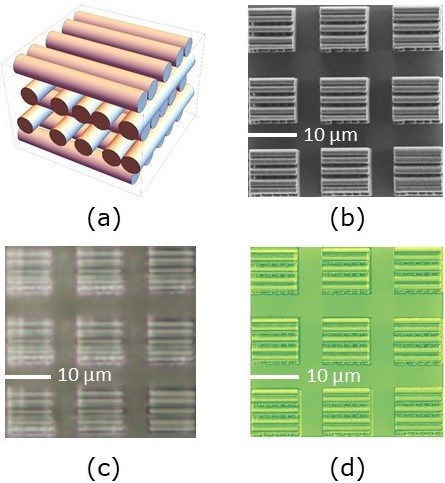}
\caption{Our samples consisting of stacks of oriented nanowires. 
(a) Sample design as a bird's eye view, showing one stack that consists of five layers of nanowires. 
In each layer, the nanowires are parallel, and have a random distance to the neighboring wire. 
(b) Scanning electron micrograph (SEM) of the sample consisting of $3 \times 3$ stacks of nanowires. 
(c) Optical microscope image of the stacks of nanowires.
(d) Overlay of sample design with scanning electron micrograph (SEM) images. 
}%%endcaption 
\label{fig:SampleDesign}
\end{figure} 
%%%%%%%%%%%%%%%%%

We have fabricated stacks of oriented nanowires that consist of $L = 3$ to $9$ layers in the $z$-direction, see the design in Figure~\ref{fig:SampleDesign}(a). 
The nanowire samples were made by two-photon polymerization in a photo-sensitive polymer (Ip-Dip, Nanoscribe GmbH) using a commercial instrument (Photonic Professional GT2, Nanoscribe GmbH). 
The nanowire stacks were fabricated on a fused silica substrate of $\mathrm{25~mm} \times \mathrm{25~mm}$ in size with a thickness of $\mathrm{0.7~mm}$. %%$0.7 mm$. 
The substrate was rinsed thoroughly using acetone and blow dried under a nitrogen flow to remove surface impurities prior to sample preparation.
The samples were prepared by drop-casting the Ip-Dip polymer on to the fused silica substrate. 
After the writing process, the sample is placed for $5-10$ minutes in a beaker containing the resist developer, subsequently rinsed with iso-propyl alcohol, and finally air-dried to obtain the desired nanowire stacks. 
The choice for a ($100\times$NA $1.4$) objective for direct laser writing was determined by the printing configuration, namely Dip-in laser lithography (DiLL). 
In this method, the microscope objective that focuses the light to induce the photopolymerization, is "dipped" into the polymer sample volume, thus producing the nanowire structures. 

Each nanowire has a diameter $d$ between $1~\mu \text{m}$ to $2~\mu \text{m}$ and a length $b = 10~\mu \text{m}$. 
Thus, the samples have physical thickness between $L = 3~\mu \text{m}$ and $20~\mu \text{m}$. 
Within each layer, all nanowires are parallel with a random nearest-neighbor spacing. 
Hence our samples are random in the $(x,y)$-directions and periodic in the $z$-direction, with orientational order within each layer. 

Each sample consists of $3 \times 3$ stacks, where the laser power was varied between the stacks in the $x$-direction and the writing speed was varied between stacks in the $y$-direction. 
We varied the writing speed between $150~\mu \text{m}$/s and $1100~\mu \text{m}$/s, and the writing laser power between $6.4$ mW and $10$ mW to search for optimal fabrication conditions. 
Increasing the laser power in the writing focus increases the diameter of the nanowires, whereas increasing the writing speed leads to deviations in conformal structure. 
Figure~\ref{fig:SampleDesign}(b) shows a scanning electron micrograph (SEM) image of a typical sample, and Figure~\ref{fig:SampleDesign}(c) an optical micrograph of the same sample. 
All 9 stacks are seen to be \textit{bona fide} realizations of the design, as shown in the overlaid optical and electron micrographs in Figure~\ref{fig:SampleDesign}(d).

%%%%%%%%%%%%%%%%%
\subsection{Optical setup}\label{sec:optical-setup}
%%%%%%%%%%%%%%%%%
\begin{figure}[htbp]
\centering\includegraphics[width=12cm]{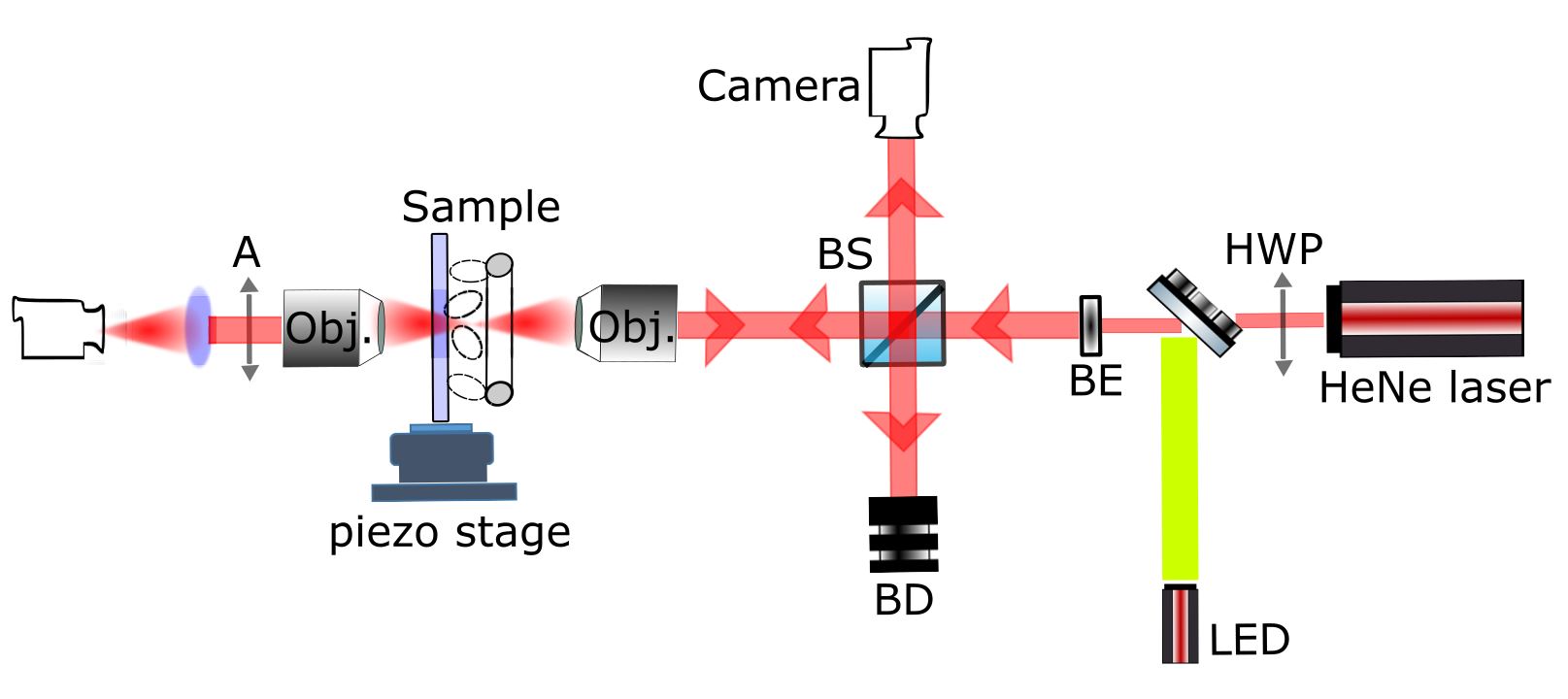}
\caption{Experimental setup to measure the total transmission. BS: beam splitter; BD: beam dump; BE: beam expander; Obj.: microscope objective (10X, 100X); HWP: half-wave plate; A: analyzing polarizer.}
\label{fig:ExperimentalSetup}
\end{figure} 
%%%%%%%%%%%%%%%%%  
Figure~\ref{fig:ExperimentalSetup} shows the experimental setup used to perform optical transmission experiments. 
We use a HeNe laser ($\lambda = 633~$nm) for transmission and an LED source to illuminate the samples for imaging. 
A beam splitter splits the illumination beam of the sample and directs the reflected light on a camera for viewing purposes. 
A flip mirror serves to choose the illumination source: 
The LED source is used to locate the nanowire stacks in reflection imaging, whereas the narrow band HeNe laser is used for total transmission and polarization-resolved transmission measurements respectively. 
A 10X objective is used to focus light onto the sample. 
To locate the nanowire stacks, the sample is mounted on a piezo stage (SmarAct MCS-3C) to finely control the (x,y) translation with a resolution of $\Delta (x,y) = 1~\mu \text{m}$.  
The transmitted light is collected by an objective (Nikon LU Plan Fluor 100$\times$/0.90 N.A.) and detected with a camera (Allied Vision, Stingray). 
For the polarization-resolved measurements, we use a half-wave plate to vary the input polarization and an analyzer to record the output polarization. 
%%%%%%%%%%%%%%%%% 
\begin{figure}[htbp]
\centering\includegraphics[width=10cm]{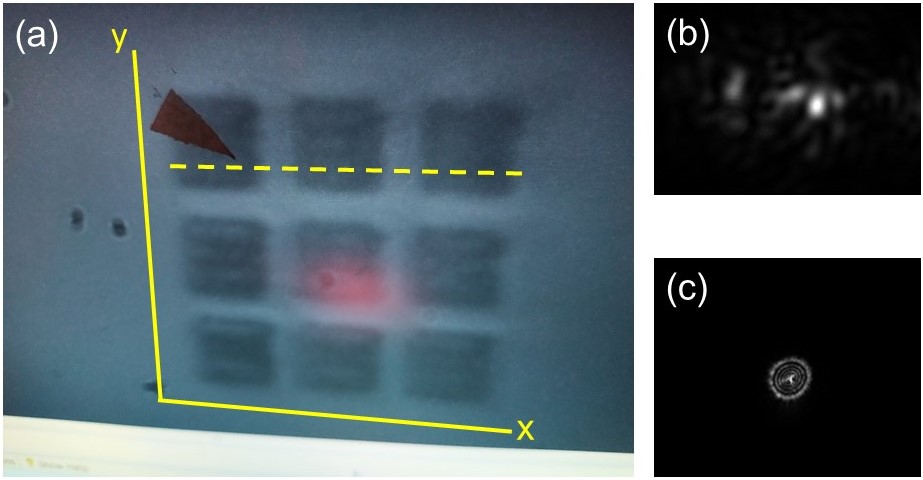}
\caption{Optical measurement. 
(a) (x,y) coordinate system of the sample and the dashed line shows how a typical position scan is done. 
(b) Reflection image captured by the camera, showing speckle. 
(c) Transmission image captured by the camera. }
\label{fig:OpticalMeasurement}
\end{figure} 
%%%%%%%%%%%%%%%%%

Figure~\ref{fig:OpticalMeasurement} illustrates the scanning technique to record the transmission and reflection data once the nanowire structures are located. 
We also define the coordinate system of the nanowire sample for the total transmission measurements. 
Light is illuminated on to the sample in a direction perpendicular to the x-y axes, see Figure~\ref{fig:OpticalMeasurement}(a). 
The sample is scanned along the x-direction in micrometer steps and the transmitted intensity is recorded by a camera, see Figure~\ref{fig:OpticalMeasurement}(c). 
The x-scan is repeated for a certain number of y-steps. 
In addition to the transmitted output, a second camera  captures the reflected light, as shown in Figure~\ref{fig:OpticalMeasurement}(b).  
%%%%%%%%%%%%%%%%%
\section{Results and Discussion}\label{sec:results}
%%%%%%%%%%%%%%%%%

%%%%%%%%%%%%%%%%%
\subsection{Total transmission}
%%%%%%%%%%%%%%%%%

%%%%%%%%%%%%%%%%%
\begin{figure}[htbp]
\centering\includegraphics[width=0.7\columnwidth]{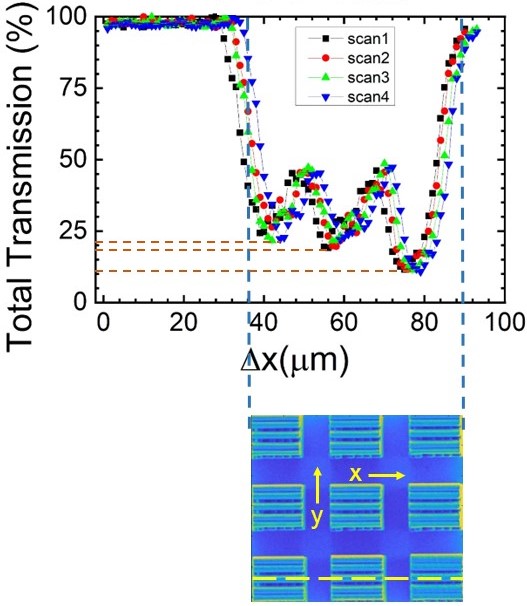}
\caption{Total transmission as a function of x-coordinate for the sample with $L = 5$ and $d = 1.5\,\mu \text{m}$. 
The scan was done four times (see the legend) through the top three structures (sample shown at the bottom), as indicated by the horizontal dashed line. 
The E-field polarization of the incident illumination is parallel to the top layer of nanowires. 
}
\label{fig:TotalTransmissionXSCan}
\end{figure} 
%%%%%%%%%%%%%%%%%
We report in Figure~\ref{fig:TotalTransmissionXSCan} total transmission measurements from the nanowire stacks while they are scanned along the x-axis.
We use the notion \textit{total transmission} (or diffuse transmission), since we employ an objective with a large numerical aperture ($NA = 0.9$) that collects a wide range of angles of the outgoing scattered light, representative of nearly all outgoing angles~\cite{vanRossum1999RMP}. 
For ease of interpretation, the transmission through the silica substrate outside the nanowire structures is set to $100 \%$ for comparison with the transmission within the structure.\footnote{With this choice, we neglect weak (Fresnel) transmission losses through the silica substrate of a few percent points.}%%endfootnote
While scanning in the x-direction, we traverse 3 different structures, as shown in Figure~\ref{fig:OpticalMeasurement}(a). 
Inside each of the 3 structures, the total transmission decreases dramatically as is apparent in Figure~\ref{fig:TotalTransmissionXSCan}, down to a minimum total transmission $T_{min} = 21 \%$, then $18 \%$ in the second structure, and finally $12 \%$ in the third structure. 
The minima decrease with increasing x, which agrees with the microscopic images that the sample on the right having thicker nanowires being naively expected to reveal stronger scattering. 
The minimum total transmission for the rightmost structure amounts to about $11 \%$, which corresponds to an estimated mean free path of about $0.94\,\mu \text{m}$, which is much less than the thickness of the sample by about a factor $3 \times$.
Therefore, we conclude that the sample is optically thick, in other words, the incident light is multiply scattered~\cite{bohren1987ajp}. 
The experiments are repeated four times and the total transmission is found to  reproduce very well, indicative of a good sample quality thanks to the sensitive fabrication equipment.

%%%%%%%%%%%%%%%%%%%%%%%%%%%%%%%%%%%%%%
\subsection{Mie scattering theory of nanowires}\label{sec:scatt-crosssection}
\begin{figure}[htbp]
\centering\includegraphics[width=\columnwidth]{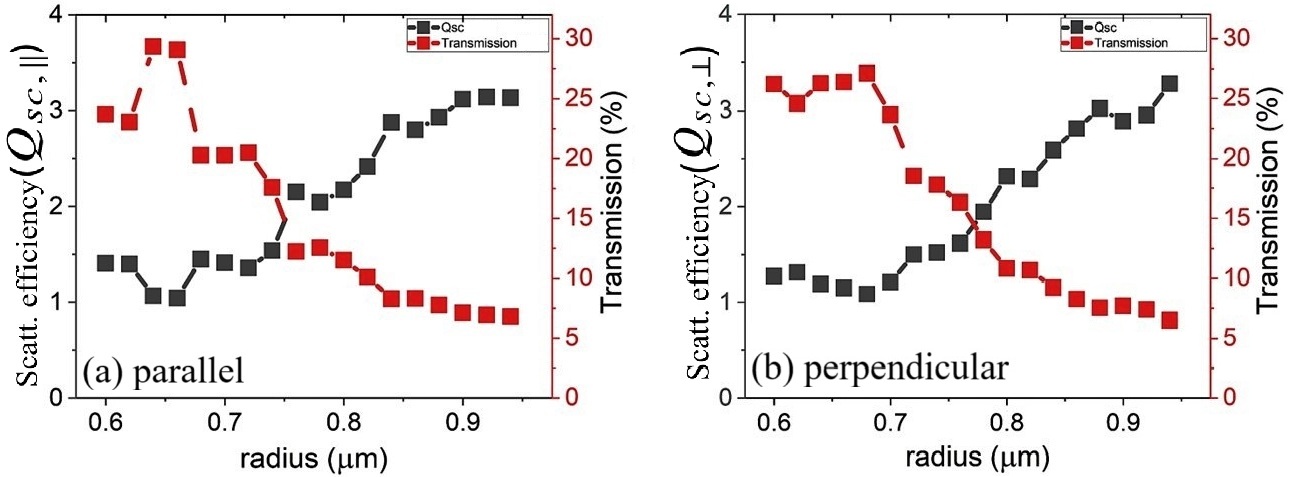}
\caption{Scattering efficiency $Q_{sc}$ of the nanowires versus radius $a$ computed with the \textbf{bhyl} program~\cite{scatterlib} for (a) parallel and (b) perpendicular light polarization at $\lambda = 633$~nm.} 
%%endcaption
\label{fig:QscTransmissionRadius}
\end{figure}

%%%%%%%%%%%%%%%%%%%%%%%%%%%%%%%%%%%%%%
To interpret the total transmission results shown in Figure~\ref{fig:TotalTransmissionXSCan} for our multi-layer nanowire stacks, we first compute the scattering cross-sections, from which we will derive the total transmission below. 
To compute the scattering cross-section of wires, we employ the Mie theory for light that is scattered by an infinite long right circular cylinder~\cite{Bohren1998book}. 
The infinite-length assumption seems reasonable since the length of the cylinders $b = 10~\mu \text{m}$ is one order of magnitude greater than their radius of $a = 0.75~\mu \text{m}$. 
For conceptual simplicity, we consider the case where the light is normally incident on the cylinder, as on the optical axis in our experiments and as discussed in detail by Bohren and Huffman~\cite{Bohren1998book}, whose \textbf{bhcyl} FORTRAN program from scatterlib\cite{scatterlib} is used here. 
%%Calculating the scattering efficiencies, that are an intensity properties, for the parallel and perpendicular polarizations, allows us to interpret the total transmission results. 
For instance, for nanowires with radius $a = 0.75~\mu \text{m}$ and the laser wavelength 633 nm, the scattering efficiency $Q_{sc}$ is equal to $Q_{sc} = 1.7898$, where we take the mean for parallel and perpendicular polarizations. 
Scattering efficiencies as a function of cylinder radii are shown for both light polarizations separately in Fig.~\ref{fig:QscTransmissionRadius}. 
For both parallel (left) and perpendicular (right) polarizations, the scattering efficiency increases with increasing radius $a$, from about $Q_{sc} = 1$ at radius $a = 0.6~\mu \text{m}$ to about $Q_{sc} = 3$ at $a = 0.9~\mu \text{m}$. 
Since the scattering efficiencies significantly exceed unity, this confirms that our nanowires are indeed in the Mie-scattering regime, and not in the Rayleigh or Rayleigh-Gans regimes~\cite{Bohren1998book}. 

%%%%%%%%%%%%%%%%%%%%%%%%%%%%%%%%%%%%%%
%%%%%%%%%%%%%%%%%%%%%%%%%%%%%%%%%%%%%%

From the scattering efficiencies we approximate the total (or diffuse) transmission $T$ as 
%%%%%%%%%%%%%%%%%%%%%%%%%%%%
\begin{equation}
\label{eq:transmission}
    T \simeq \frac{1}{\rho C_{sc}} = \frac{1}{\rho~Q_{sc}~G}, 
\end{equation}
%%%%%%%%%%%%%%%%%%%%%%%%%%%%
where $\rho$ is the number density of the nanowires, $C_{sc}$ the scattering cross section (units: area), and $G$ the geometrical cross section (equal to $G = 2~a~b$), where we use the notation of Bohren and Huffman~\cite{Bohren1998book}.
In Eq.~\ref{eq:transmission}, we neglect internal reflection boundary effects. 
Moreover, we employ the independent scattering approximation (ISA)~\cite{vanRossum1999RMP}.
Figure~\ref{fig:QscTransmissionRadius} shows the total transmission $T$ for parallel (left) and perpendicular (right) polarizations, as a function of nanowire radius $a$. 
The total transmission $T$ decreases from about $25 \%$ to $7 \%$, for both polarizations. 
By estimating the corresponding absolute scattering cross-section $C_{sc}$ and density $\rho$ using the known nanowire radii $a$ and lengths $b$, we derive that the mean free path for nanowires with an examplary radius $a = 0.75~\mu \text{m}$ is equal to $\ell= 1.12~\mu \text{m}$. 
The mean free path allows us to estimate the total transmission to be $T = \frac{\ell}{L} = \frac{1.12}{7.5} = 15\%$, which agrees well with our observations for the largest nanowire structure of $T = 12\%$, that is shown in Figure~\ref{fig:TotalTransmissionXSCan}. 

%%%%%%%%%%%%%%%%%%%%%%%%%%%%%%%%
\begin{table}[htbp]
\begin{tabular}{|c|c|c|c|c|}
  \hline
 Structure  &  $a(\mu \text{m})$ from ${Q_{sc, \perp}}$ & $a(\mu \text{m})$ from $Q_{sc, \parallel}$ & $a_{Mean}(\mu \text{m})$ & $a_{SEM}(\mu \text{m})$ \\
  \hline
  1 & $0.679$  & $0.662$ & $0.670 \pm 0.009$ & $0.795 \pm 0.066$  \\
  2 & $0.698$  & $0.675$ & $0.686 \pm 0.012$ & $0.840 \pm 0.066$  \\
  3 & $0.780$  & $0.756$ & $0.768 \pm 0.012$ & $0.883 \pm 0.066$  \\
  \hline
\end{tabular}
%%%%%%%%%%%%%%%%%%%%%%%%%%%%%%%%
\caption{List of nanowire samples with radii $a$ estimated from the scattering efficiencies for both polarizations ($\perp$ and $\parallel$), their mean optical radii, and the radii $a_{SEM}$ obtained from the SEM images. }%%endcap
\label{table:ScattSEMRadiiTable}
\end{table}
%%%%%%%%%%%%%%%%%%%%%%%%%%%%%%%%

From the observed total transmissions shown in Figure~\ref{fig:TotalTransmissionXSCan} and the comparison with the calculated scattering efficiencies shown in Figure~\ref{fig:QscTransmissionRadius}, we infer the nanowires to have radii as listed in Table~\ref{table:ScattSEMRadiiTable}. 
Table~\ref{table:ScattSEMRadiiTable} also lists radii obtained from scanning electron microscopy, shown in Figure~\ref{fig:SampleDesign}(b). 
Overall, the radii estimated from the optical total transmission and scattering efficiencies agree well with those derived from electron microscopy. 
The optical estimates are systematically about $10 \%$ smaller than the SEM results, which we tentatively attribute to our neglect of the internal optical reflection of our scattering samples~\cite{Lagendijk1988PhysLett, Zhu1991PRA}. 
For the third structure, the mean radii are found to be $a = 0.77\,\mu \text{m}$, which is close to the designed radius $a = 0.75\,\mu \text{m}$. 
From the good agreements, we conclude that our optical model is a faithful description of our results, and hence that the mean free path is substantially smaller than the sample thickness $L$. 
Therefore, we draw the overall conclusion that our nanowire samples are optically thick, in other words, the samples are in the multiple scattering regime~\cite{vanRossum1999RMP, Carminati2021book}.
%%%%%%%%%%%%%%%%%%%%%%%%%%%%%%%%%%%%%%%%%%%
\subsection{Polarized multiply scattered light}\label{sec:polarized}
%%%%%%%%%%%%%%%%%%%%%%%%%%%%%%%%%%%%%%
%%%%%%%%%%%%%%%%%%%%%%%%%%%%%%%%%%%%%%
\begin{figure}[htbp]
\centering\includegraphics[width=0.95\columnwidth]{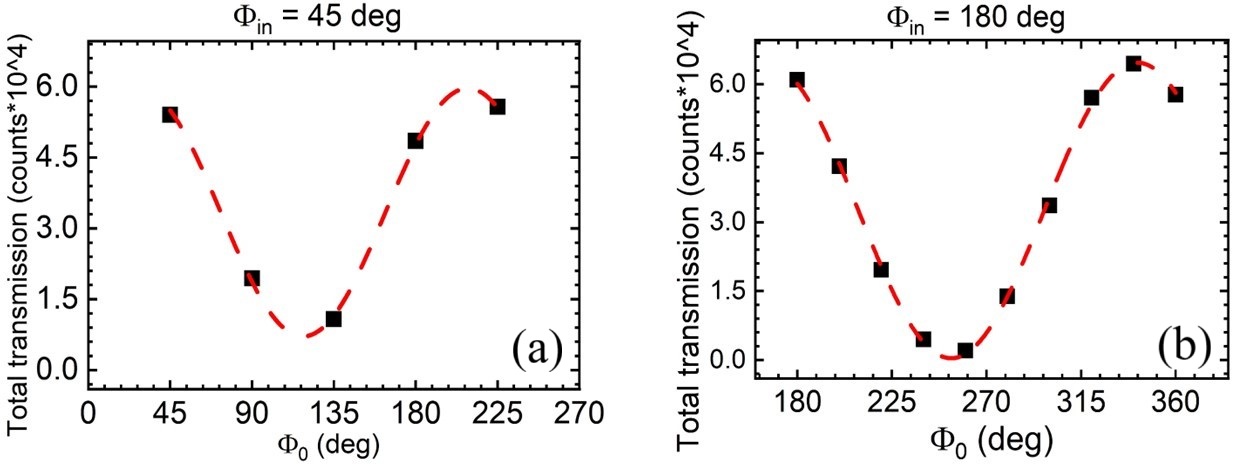}
\caption{Total transmission measured as a function of outgoing scattered polarization (black squares) for a sample with thickness $L = 5\,\mu \text{m}$ and $a = 0.75\,\mu \text{m}$ for two different input polarizations: 
(a) $\Phi_{in} = 45^\circ$, 
(b) $\Phi_{in} = 180^\circ$. 
The red dashed curves are sinusoidal models of the measured data.}%%endcaption
\label{fig:TotalTransmissionPolarization1}
\end{figure} 
%%%%%%%%%%%%%%%%%%%%%%%%%%%%%%%%%%%%%%
So far our analyses were restricted to two limiting input polarization angles $0^\circ$ and $90^\circ$ corresponding to light polarizations parallel and perpendicular to the nanowires. 
We now turn to the total transmission as a function of both input and output polarizations and include input polarizations ranging from $0^\circ$ to $180^\circ$ in $45^\circ$ steps, where we separately analyze the output polarization to investigate the nanowire scattering response to polarized radiation. 
Figure~\ref{fig:TotalTransmissionPolarization1} shows total transmission for two different input polarizations $\Phi_{in} = 45^\circ$ (left) and $\Phi_{in} = 180^\circ$ (right), for a nanowire sample with thickness $L = 5\,\mu \text{m}$ and nanowire radius $a = 0.75\,\mu \text{m}$, respectively. 
The output analyzer polarization $\Phi_\mathrm{o}$ was varied in several steps and the total transmission recorded, as shown in Figure~\ref{fig:TotalTransmissionPolarization1}. 
It is seen that the transmitted \textit{diffuse} intensity varies strongly with \textit{output} polarization, with variations ranging from maxima of about $6 \times 10^{+4}$ counts to minima less than $0.5 \times 10^{+4}$ counts, in other words, more than $10\times$ variations with output polarization, whereas the input polarization and sample orientation are constant. 
Furthermore, the output polarization $\Phi_\mathrm{o}$ that corresponds to the maximum transmission tracks the input polarization $\Phi_{in}$ quite closely, as a linear dependence. 
Indeed, the observed transmission data match very well to a sinusoid, from which we determine the output polarization angles corresponding to maximum and minimum transmission, respectively.

%%%%%%%%%%%%%%%%%%%%%%%%%%%%%%%%%%%%%%
\begin{figure}[htbp]
\centering\includegraphics[width=7cm]{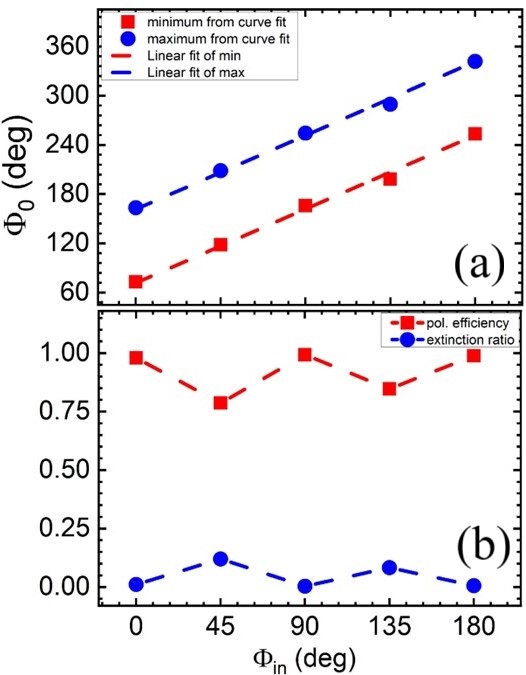}
\caption{Polarization parameters using Maxima and minima values of $\Phi_{0}$. (a) absolute $\Phi_\mathrm{o}$ maxima and minima from curve fit in Fig.~\ref{fig:TotalTransmissionPolarization1}. (b) polarization efficiency and extinction ratio calculated from using integrated intensity at $\Phi_\mathrm{o}$ maxima and minima.}%%endcaption
\label{fig:TotalTransmissionPolarization2}
\end{figure} 
%%%%%%%%%%%%%%%%%
Figure~\ref{fig:TotalTransmissionPolarization2}\,(a) shows the output polarization angles as obtained from sinusoidal models versus the input polarization angle. 
The linear tracking of output polarization $\Phi_\mathrm{o}$ with input polarization $\Phi_{in}$ is readily apparent from these data. 
Thus, depite the fact that the light is multiply scattered by the nanowires (see section~\ref{sec:scatt-crosssection}), the polarization of the scattered light is \textit{not} scrambled by the nanowires. 
%%%%%%%%%%%%%%%%%%%%%%%%%%%%%%%%%%%%%%%%%%%%%%%%%%%%%%%%%%%%%%%%%%%
\subsection{Interpretation and discussion }\label{sec:interpretation}
%%%%%%%%%%%%%%%%%%%%%%%%%%%%%%%%%%%%%%
To summarize our essential results at this point, we observe that samples of oriented perpendicular cilindrical scatterers (nanowires) are on one hand optical thick (or multiple scattering) when we consider the transported intensity of the light, whereas simultaneously the samples are transparent for the polarization of the light. 
The latter feature is at odds with common lore in optically thick samples composed of spherical scatterers where polarization is scrambled to random output polarization states~\cite{vanRossum1999RMP, Carminati2021book}. 
Here, in contrast, we observe that oriented nanowire samples do not scramble the input polarization at all.
To interpret these features, we invoke the discussion on light scattering off cylinders by Bohren and Huffman (see their Chapter 8)~\cite{Bohren1998book}. 

%%%%%%%%%%%%%%%%%%%%%%%%%%%%%%%%%%%%%%
\begin{figure}[htbp]
\centering\includegraphics[width=0.9\columnwidth]{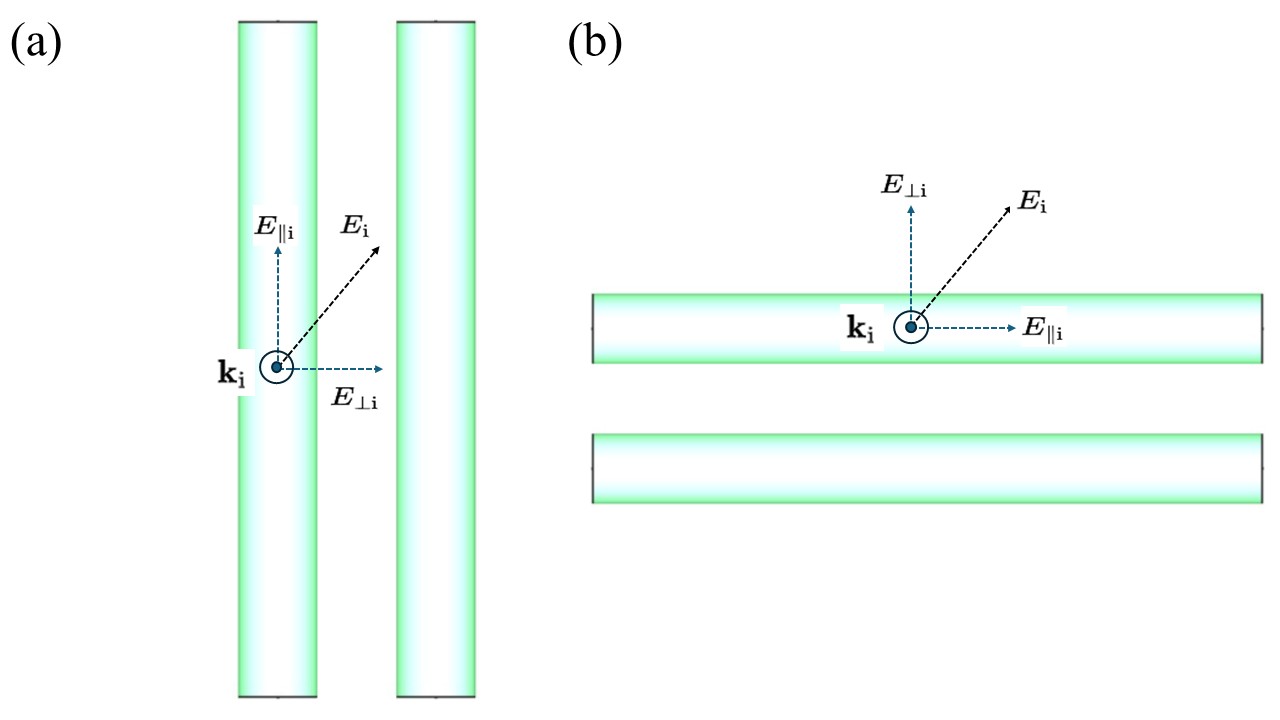}
\caption{Cartoons of polarized light incident on nanowires in one layer, taken to have an incident wave vector $\mathbf{k}_\mathrm{i}$ into the plane. 
(a) A polarization vector $\mathbf{E}_{i}$ (drawn) is inclined compared to the vertical nanowires, and is decomposed in parallel ($E_{\parallel \mathrm{i}}$) and perpendicular ($E_{\perp \mathrm{i}}$) components.
(b) The same polarization $\mathbf{E}_{i}$ is incident onto horizontally oriented nanowires ($E_{\parallel \mathrm{i}}$ and $E_{\perp \mathrm{i}}$ swapped compared to (a)). 
}%%endcaption
\label{fig:cartoonpolarization}
\end{figure} 
%%%%%%%%%%%%%%%%%

In our discussion, we firstly assume that light is incident normal to the cylinders or nanowires. 
This is a simplification of our experimental situation, where light is focused in a cone onto the sample, but is useful for the sake or reasoning. 
Moreover, if incident light has a wave vector $\mathbf{k}_\mathrm{i}$ that is inclined with the nanowire axis, it is reasonable that only the component perpendicular to the nanowire $\mathbf{k}_{\perp \mathrm{i}}$ will be scattered, whereas light with a parallel component $\mathbf{k}_{\parallel \mathrm{i}}$ will hardly be scattered since the nanowires are homogeneous along their length. 
Furthermore, it may be reasoned that our samples are so dense that they have an effective refractive index of about $n_\mathrm{eff} = 1.4$, hence light is refracted towards the normal, so a sizeable cone of incident directions is effectively nearly perpendicular to the plane of the nanowires, where this discussion pertains to.\footnote{Our illumination objective has $\mathrm{NA}= 0.25$, hence a maximum external incidence angle $\theta_\mathrm{out} = 14.5^{\circ}$, a corresponding maximum internal incidence $\theta_\mathrm{in} = 10^{\circ}$ (hence $\mathrm{cos}(\theta_\mathrm{in}) = 0.98$), thus the internal wave vector is mostly perpendicular to the nanowires axes.} 

As illustrated in Figure~\ref{fig:cartoonpolarization}, the polarization vector $\mathbf{E}_{i}$ is readily decomposed in components parallel ($E_{\parallel \mathrm{i}}$) and perpendicular ($E_{\perp \mathrm{i}}$) to a nanowire. 
From Bohren and Huffman's equation (8.40)~\cite{Bohren1998book}, reproduced here: 
%%%%%%%%%%%%%%%%%%%%%%%%%%%%%%%%%%%%%%
\begin{equation}
\begin{pmatrix}
E_{\parallel \mathrm{o}} \\
E_{\perp \mathrm{s}}
\end{pmatrix}
= e^{i 3\pi / 4} \sqrt{\frac{2}{\pi k r}}\, e^{i k r} 
\begin{pmatrix}
T_{1} & 0 \\
0 & T_{2}
\end{pmatrix}
\begin{pmatrix}
E_{\parallel \mathrm{i}} \\
E_{\perp \mathrm{i}}
\end{pmatrix}, 
\label{eq:polarization_cylinder}
\end{equation}
%%%%%%%%%%%%%%%%%%%%%%%%%%%%%%%%%%%%%%
with $k$ the modulus of the wave vector, $r$ the distance to the detector, and $T_{1}$ and $T_{2}$ scattering amplitudes. 
We see in Eq.~\ref{eq:polarization_cylinder} that the scattered output field components $E_{\parallel \mathrm{o}}$ and $E_{\perp \mathrm{o}}$ have the same phase shifts compared to the incident field components $E_{\parallel \mathrm{i}}$ and $E_{\perp \mathrm{i}}$.
Therefore, a linear polarization, as illustrated in Fig.~\ref{fig:cartoonpolarization}, will be scattered into a linearly polarized output, whereby unequal scattering amplitudes ($T_{1} \neq T_{2}$) describe a rotation of the output polarization compared to the incident polarization. 
Therefore, even after scattering by multiple layers of parallel nanowires, the multiple scattered polarization remains linear, in agreement with our observations. 

Let us zoom in on the scattering cross-sections, as shown in Figure~\ref{fig:QscTransmissionRadius} and in Figure~8.7 of Bohren and Huffman~\cite{Bohren1998book}. 
For cylinder diameters beyond $1.0~\mu \text{m}$, corresponding to $a \geq 0.5~\mu \text{m}$, as in our study, it is apparent that the scattering cross sections for light parallel and perpendicular to the cylinder or nanowire axis are equal to each other. 
Therefore, it is reasonable to assume in the discussion above that the two scattering amplitudes are nearly the same: $T_{1} \simeq T_{2}$.
Thus, the output polarization vector is oriented parallel to the input polarization vector. 
From this reasoning, we arrive at the exhilarating conclusion that when light is scattered by a stack of nanowires, the output polarization vector remains parallel to the input polarization, which is indeed the ``polarization maintaining multiple scattering'' apparent in Figure~\ref{fig:TotalTransmissionPolarization2}. 
Our discussion on polarization maintaining multiple scattering may be readily generalized to other classes of anisotropic light scattering samples beyond the polymer nanowires studied here. 
The essence is that the discussion pertains to samples that consist of many anisotropic scatterers that are arranged in layers. 
Thus, our discussion pertains to all structures that consist of high-index wire-like scatterers in a low-index backbone, which includes polymer wires as studied here, as well as samples consisting of oxide (nano)wires~\cite{Dattoli2011}, and of semiconductor (nano)wires or (nano)rods~\cite{Strudley2014ol,Burke2025APL,Iwamoto2016Ph}.
Moreover, since Bohren and Huffman's scattering equation (8.40) is valid for a cylinder without restriction on the dielectric function, it also pertains to inverse structures consisting of low-index wire-like scatterers in a high-index backbone. 
Hence, our discussion also pertains to pores that are etched in silicon and other semiconductors, including popular woodpile and inverse woodpile photonic crystals, as studied by our group and elsewhere~\cite{Lin1998nat, Ogawa2004sci, Garcia-Santamaria2007AdvMa, Hermatschweiler2007AdvMat, vandenBroek2012AdvFuncMat, Marichi2016SciRep, Vreman2025prb}. 
Our discussion also pertains to more general structures that consist of layers of anisotropic nanowire scatterers, when the adjacent layers of scatterers are not perpendicular (as in Figs.~\ref{fig:SampleDesign} and~\ref{fig:cartoonpolarization}), but subtend any possible angle. 
Moreover, subsequent layers do not have to be parallel to each other, so the structures may even be a random stacking of layers. 
These structures are amenable to our discussion, as it only invokes polarized scattered by a single layer of anisotropic scatterers.  
%%%%%%%%%%%%%%%%%%%%%%%%%%%%%%%%%%%%%%%%%%%%%%%%%%%%%%%%%%%%%%%%%%%
\subsection{Possible device applications}\label{sec:application}
%%%%%%%%%%%%%%%%%%%%%%%%%%%%%%%%%%%%%%
Let us briefly discuss several possible applications of the class of anisotropic samples studied here, and notably the peculiar polarization behavior where the sample is basically transparent to the polarization of light. 
Light transport in multiply scattering media is of great interest in various fields of research and industry, and polarized light scattering is employed in bio-medical imaging, lighting applications, and light harvesting systems and solar cells~\cite{deBoer2002JBO, Kienle2004OptLet, Judkewitz2015NatPhys, Antony2023PPS}.
The wide proliferation arises because controlled transmission of photonic polarization states can readily be used to re-direct light more efficiently between scatterers. 
A major class of practical optical devices are white light emitting devices (LEDs) that are produced by the lighting industry~\cite{Schubert2006Book, Amano2015RMP, Akasaki2015RMP, Nakamura2015RMP}, with an economic value of tens of billions of euros, for applications in lighting - at home, in buildings, in public spaces - and myriad other functionalities ranging from ``light fidelity'' (LiFi) communication via encryption~\cite{Haas2015JLT, Rates2023OptEx}, to biomedical sensing with miniature handheld and point-of-care (POC) devices~\cite{Bogomolov2017sensors,keller2018boe}. 
Many modern LEDs consist of a blue diode emitter on top of a phosphor that serves to convert part of the blue to other colors (green, yellow, orange, red) for appropriate color rendering, and to scatter the light to yield a desired diffuse output~\cite{Schubert2006Book}. 
Typically, the output light is unpolarized due to the intrinsic scattering properties of the phosphor. 
Here, we envision the incorporation of anisotropic nanowires as light scatterers elements in the phosphor. 
Whereas the converted colors will remain unpolarized (since they are emitted with random polarizations), the incident blue light may retain its original polarization, for possible secondary applications of sensing, or signaling, or communication. 
This thus allows additional functions for a white LED without additional (bulky, complex, or expensive) optical components and make lighting systems more versatile. 
In the interest of large scale fabrication, we surmise that the nanowires may be made by (bulk) chemical means, and be deposited on suitable substrates by liquid flow or external electric fields. 
The transmitted light may be further controlled by optimizing individual nanowire and layer dimensions such as the diameter of a nanowire and the number of stacked layers.

%%%%%%%%%%%%%%%%%%%%%%%%%%%%%%%%%%%%%%
\section{Summary}\label{sec:summary} 
%%%%%%%%%%%%%%%%%%%%%%%%%%%%%%%%%%%%%%
In this paper, we study the transport of light through flat and dense anisotropic samples of oriented nanowires. 
The nanowires are made from a polymer photoresist by direct laser writing (DLW) with two-photon induced polymerization, with radii ranging from $a = 0.5$ to $1~\mu \text{m}$. 
Our thin samples have a thickness of up to 9 layers, corresponding to $L = 3~\mu \text{m}$ to $20~\mu \text{m}$, and extend over an area of $10~\times 10~\mu \text{m}^{2}$.   
Within each layer, the nanowires have random nearest-neighbor distances, and are oriented parallel to each other, and perpendicular to nanowires in adjacent layers. 
The diffuse optical transmission of the samples at $\lambda = 633~$nm is as low as $T = 12 \%$, characteristic of a fairly strongly scattering sample with a mean free path of the order of $\ell = 1.1~\mu \text{m}$ that is clearly less than the sample thickness. 
It is a striking feature that when the incident polarization is rotated, the outgoing polarization closely tracks the incident one and remains remarkably linear. 
Conversely, the outgoing polarization is hardly determined by the nanowire orientation, even though the light is multiply scattered. 
We discuss that the polarization is remarkably maintained in our optically thick samples, since the polarization incident on each layer of nanowires is faithfully described with parallel and perpendicular components; moreover, the components incident from a previous layer are also well described as parallel and perpendicular in the next layer, since the nanowires are exactly perpendicular to those in the previous layer. 
The polarization is maintained in our optically thick samples, likely since the light is mostly transported in the z-direction perpendicular to the nanowires. 
We discuss applications of the phenomena in white LEDs for lighting and additional sensing and signaling purposes. 

%%%%%%%%%%%%%%%%%%%%%%%%%%%%%%%%%%%%%%
\section*{Acknowledgments} 
%%%%%%%%%%%%%%%%%%%%%%%%%%%%%%%%%%%%%%
We thank Cornelis Harteveld for technical support and Nicole Meinster for administrative support, the MESA+ Nanolab staff for help with the nanofabrication, and Wilbert IJzerman and Gilles Vissenberg (Signify), Teus Tukker (ASML), Bart van Tiggelen (LPMMC, Grenoble), Leon Woldering (Demcon), Ozan Akdemir, Ad Lagendijk, and Pepijn Pinkse for helpful discussions. 
This work was supported by the NWO-TTW Perspectief program P15-36 ‘Free-form scattering optics’ (FFSO) in collaboration with TU/e and TUD and with industrial partners ASML, Demcon, Lumileds, Schott, Signify, and TNO, by the ``Descartes-Huygens" prize of the French Academy of Sciences. 
We dedicate this paper to Bart van Tiggelen, a great and kind theorist who inspired much research on anisotropic light scattering. 

%%%%%%%%%%%%%%%%%%%%%%%%%%%%%%%%%%%%%%

\bibliographystyle{apsrev4-2}
\bibliography{scattering_oriented_arxiv}

%%%%%%%%%%%%%%%%%%%%%
\end{document}